\documentclass[traditabstract]{aa} 
\usepackage{graphicx}
\usepackage[sectionbib]{natbib}
\usepackage{txfonts}
\begin{document}
\title{Fragmentation of a dynamically condensing radiative layer}

   \author{Kazunari Iwasaki
          \and
          Toru Tsuribe 
          }

   \institute{Department of Earth and Space Science, Graduate School
              of Science, Osaka University, Toyonaka, Osaka 560-0043, Japan
              \\
              \email{[iwasaki; tsuribe]@vega.ess.sci.osaka-u.ac.jp}
             }

   \date{Received ; accepted }

 
  \abstract
  {
  In this paper, 
  the stability of a dynamically condensing radiative gas layer is investigated by linear 
  analysis.  
  Our own time-dependent, self-similar solutions describing a dynamical condensing 
  radiative gas layer are used as an unperturbed state.
  We consider perturbations that are both perpendicular and parallel 
  to the direction of condensation.
  The transverse wave number of the perturbation is defined by $k$.
  For $k=0$, it is found that the condensing gas layer is unstable. However, 
  the growth rate is too low to become nonlinear during dynamical condensation.
  For $k\ne0$, in general,
  perturbation equations for constant wave number cannot be reduced to 
  an eigenvalue problem 
  due to the unsteady unperturbed state.
  Therefore, direct numerical integration of the perturbation equations is performed. 
  For comparison, an eigenvalue problem neglecting the time evolution 
  of the unperturbed state
  is also solved and both results agree well.
  The gas layer is unstable for all wave numbers, and the growth rate depends a little
  on wave number.
  The behaviour of the perturbation is specified by $kL_\mathrm{cool}$ at the centre, where 
  the cooling length, $L_\mathrm{cool}$, represents the length that a sound wave can travel 
  during the cooling time. For $kL_\mathrm{cool}\gg1$, 
  the perturbation grows isobarically.
  For $kL_\mathrm{cool}\ll1$, 
  the perturbation grows because each part has a different collapse time without interaction.
  Since the growth rate is sufficiently high, it is not long before the perturbations become nonlinear
  during the dynamical condensation. Therefore, according to the linear analysis, 
  the cooling layer is expected to 
  split into fragments with various scales.
  }


   \maketitle
%

\section{Introduction}
In the interstellar medium (ISM), it is well known that 
a clumpy low-temperature phase (cold neutral medium, or CNM) 
and a diffuse high-temperature phase (warm neutral medium, or WNM) can coexist in 
pressure equilibrium as a result of the balance of 
radiative cooling and heating due to external radiation
fields and cosmic rays \citep{FGH69,W95,W03}. 
These two phases are thermally stable. On the other hand,
gas is thermally unstable in the temperature range between 
two stable phases, that is, in the range 300 K $<T<$ 6000 K. 
The unstable gas spontaneously turns into the mixture of stable CNM and WNM by 
thermal instability (TI).
This instability is expected to play an important role 
in the structure formation and the dynamics of the ISM,
and especially, in the molecular cloud formation and origin of turbulence.

The basic properties of TI was investigated by \citet{F65}, who performed linear analysis
of an uniform gas in thermally equilibrium.
He derived a criterion for TI.
Focusing on one fluid element, \citet{B86} 
generalized the Field criterion 
when the cooling rate is not equal to 
the heating rate.
The effect of magnetic field on TI has been investigated by
\citet{F65} and \citet{HP06}, and other authors.

Recently, many authors have used multi-dimensional numerical simulations  
to study the turbulent CNM formation driven by TI. 
\citet{KI02} suggest that the turbulent CNM formation
is induced by TI in a shock-compressed region.
Analogous processes have been studied by many authors for a 
colliding flow of the WNM \citep{AH05, HA07, H06,V07}, 
and using two-fluid MHD simulation \citep{II08}.
The unbalance between cooling and heating rates causes the shock-compressed 
gas layer to cool and to condense.
During the cooling, these numerical simulations shows that 
the runaway cooling layer quickly fragments into many CNM clumps whose
velocity dispersion is equal to a fraction of the sound speed of WNM, 
where CNM clumps and WNM are tightly interwoven.
This complex structure is regarded as produced by TI and possibly by
some other hydrodynamical instabilities, 
such as the nonlinear thin shell instability \citep{V94},
the Kelven-Helmholtz instability, and by corrugation instability
of the phase transition layers between CNM and WNM \citep{IIK06}.

A fluid element that is compressed by a shock wave 
tends to be a layer rather than a sphere because it is only compressed 
in the direction perpendicular to the shock front.
Once the fluid element enters the thermally unstable regime, 
the layer cools in a runaway fashion.
In this paper, we focus on the fragmentation of 
the runaway cooling layer.
In previous studies, A detailed physical mechanism of the fragmentation 
of the runaway cooling layer
remains poorly understood even in linear regime.
The main reason is that
it is difficult to select the unperturbed state since the cooling layer 
evolves temporarily and spatially. Therefore, in previous works, 
the unperturbed states were limited to 
spatially uniform gas that cools isochorically 
\citep{SMS72,BL00} or isobarically \citep{KI00}.

\citet{IT08} (hereafter IT08) have recently found a family of self-similar (S-S) solutions
describing the dynamical condensation of a radiative gas layer 
where the cooling rate dominates the heating rate.
This S-S solution assumed that the net cooling rate is a power-law function and that the
heating rate is explicitly neglected. Although it is still too ideal,
they are expected to be a good 
nonlinear one-dimensional model at least in the phase during the transition from
WNM to CNM.
In this paper, we adopt the S-S solutions as a more 
realistic unsteady unperturbed state than those in previous works.
We perform linear analysis of the S-S solutions against 
fluctuations perpendicular, as well as parallel to, the direction of 
the condensation.
By performing the linear analysis, we will have some useful insights
when and how the cooling layer fragments.
Since we focus on the above S-S unperturbed state, 
the nonlinear thin shell instability, Kelvin-Helmholtz instability,
and the corrugation instability are beyond the scope of this paper. 

In Sect. \ref{formulation}, we formulate basic equations using 
a zooming coordinate.
Perturbation equations are derived
for the linear analysis, with a brief review of the S-S solutions.
In Sect. \ref{keq0}, we investigate the stability of the S-S solutions taking only those fluctuations 
into account that are parallel to the direction of the condensation.
In Sect. \ref{sec kne0}, we consider the fluctuations that are both perpendicular 
and parallel to the direction of the condensation.
In Sect. \ref{discuss}, we discuss the astrophysical implication of the linear analysis 
and effects of the thermal conduction.
Our study is summarized in Sect. \ref{summary}.
\section{Formulation}\label{formulation}
We consider a dynamically condensing radiative gas layer
where the cooling rate dominates the heating rate.
The following formula is adopted as 
the net cooling rate per unit volume and time:
\begin{equation}
        \rho{\cal L}(\rho,P) = \Lambda_0\rho^{2-\alpha}P^{\alpha-1}\propto 
        \rho^2 T^\alpha\;\;\mathrm{erg\;cm^{-3}\;s^{-1}}.
        \label{cooling func}
\end{equation}
In ISM, in the temperature range of $T\la 10^7$, the main coolant is Bremsstralung 
for the solar metallicity \citep{SD93}. 
The cooling rate of Bremsstralung is approximated well by that with $\alpha=0.5$.
In the temperature range of $2\times10^5\;\mathrm{K}\la T \la 10^7\;\mathrm{K}$,
the dominant coolant is metal lines for the solar metallicity
\citep{SD93}. The cooling rate of metal lines corresponds to that with $\alpha<0$. 
However, the cooling rate cannot be expressed by a single $\alpha$.
In the temperature range between CNM and WNM, $300\la T\la 6000$, the dominant coolant
is CII \citep{DM72}. 
In this case, the cooling rate is approximated by that with 
$\alpha\simeq0.6$ as is shown in Sect. \ref{astro impli}.

Basic equations for a radiative gas are
the continuity equation,
\begin{equation}
        \frac{\partial \rho}{\partial t} + \vec{\nabla}\cdot(\rho \vec{v})=0,
\label{eoc}
\end{equation}
the equation of motion,
\begin{equation}
        \frac{\mathrm{D} \vec{v}}{\mathrm{D} t}
        +\frac{1}{\rho}\vec{\nabla} P=0,
\label{eom}
\end{equation}
and the entropy equation,
\begin{equation}
        \frac{1}{\gamma-1}\frac{\mathrm{D} }{\mathrm{D} t}\left( \ln P \rho^{-\gamma}\right)
= - \gamma^{\alpha}\Lambda_0\rho^{2-\alpha} P^{\alpha-1},
\label{eoe}
\end{equation}
where $\mathrm{D}/\mathrm{D} t=\partial/\partial t+\vec{v}\cdot\vec{\nabla}$ 
indicates the Lagrangian time derivative.

We take the $x$-axis as the direction of 
the condensation driven by the cooling and 
$y$-axis as the transverse direction.
Since the S-S solutions are time-dependent, it
is difficult to perform linear analysis 
in the ordinary Cartesian coordinate, $(t,x,y)$.
\cite{B85} introduced a zooming coordinate 
where S-S solutions appear to be stationary 
\citep[also see][]{HM99}.
We introduce the similar zooming coordinate
since this transformation makes stability analysis easier as follows:
\begin{equation}
\left(
 \begin{array}{c}
    t \\ 
    x \\ 
    y \\
 \end{array}
\right) \Rightarrow
\left(
 \begin{array}{c}
    \tau \\ 
    \xi \\ 
    y \\
 \end{array}
\right) =
\left(
 \begin{array}{c}
         \displaystyle -\frac{\ln t_*}{1-\omega} \\ 
         x/x_0(t) \\ 
         y \\
 \end{array}
\right),
\;\; x_0(t)=at_*^{1/(1-\omega)},
\label{zooming}
\end{equation}
where $\omega$ is a free parameter, $a$ corresponds to the cooling length,
and $t_* = 1-t/t_\mathrm{c}$, $t_\mathrm{c}$ is an epoch when 
the central density becomes infinity.
In the zooming coordinate, 
density $\Omega$, velocity $\vec{V}$, pressure $\Pi$, and sound speed $X$ are given by 
\begin{eqnarray}
\Omega(\tau,\xi,y) = \rho(t,x,y)/\rho_0(t), 
&& \vec{V}(\tau,\xi,y)=\vec{v}(t,x,y)/v_0(t), \nonumber \\
\Pi(\tau,\xi,y) = P(t,x,y)/P_0(t), 
&& X(\tau,\xi,y) = c_\mathrm{s}(t,x,y)/v_0(t), 
\label{per vari}
\end{eqnarray}
respectively, where
\begin{equation}
        v_0(t)= -\dot{x}_0(t)=\frac{a}{(1-\omega)t_\mathrm{c}}t_*^{\omega/(1-\omega)},
\label{self vari1}
\end{equation}
\begin{equation}
        \rho_0(t) = \left\{ \frac{1}{(1-\omega)t_\mathrm{c}} \right\}^{-2\alpha+3}
\frac{a^{2(1-\alpha)}}{\Lambda_0}t_*^{\beta/(1-\omega)},\nonumber
\label{self vari2}
\end{equation}
and
\begin{equation}
        P_0(t) = \frac{\rho_0v_0^2}{\gamma}\propto t_*^{(2\omega+\beta)/(1-\omega)},
\label{self vari3}
\end{equation}
with $\beta=\omega(3-2\alpha)-1$.

In the zooming coordinate, 
the basic equations (\ref{eoc})-(\ref{eoe}) are rewritten as
\begin{equation}
        \frac{\mathrm{D} \ln \Omega}{\mathrm{D} \tau} 
        + \vec{\nabla}\cdot\vec{V} = \beta,
\label{sim1}
\end{equation}
\begin{equation}
        \frac{\mathrm{D} \vec{V}}{\mathrm{D} \tau} 
        +\frac{1}{\Omega}\vec{\nabla} \Pi = \omega \vec{V},
\label{sim2}
\end{equation}
and 
\begin{equation}
\frac{1}{\gamma-1}\frac{\mathrm{D} }{\mathrm{D} \tau} 
\left( \ln \Pi \Omega^{-\gamma} \right)
= \frac{2\omega}{\gamma-1} -\beta - \gamma^\alpha\Omega^{2-\alpha}\Pi^{\alpha-1},
\label{sim3}
\end{equation}
respectively, where the operators of time and spatial derivative are
defined by 
\begin{equation}
        \frac{\mathrm{D}}{\mathrm{D}\tau} = 
        \frac{\partial}{\partial \tau}
        + (\vec{V} + \xi\vec{e}_\xi)\cdot \vec{\nabla},
        \;\;\;\mathrm{and}\;\;
        \vec{\nabla} = \left( \frac{\partial}{\partial \xi}, 
        x_0(\tau)\frac{\partial}{\partial y} \right),\;\;
        \label{def ope}
\end{equation}
respectively, where $\vec{e}_\xi$ 
indicates the unit vector parallel to $\xi$-direction.
Supplements for the derivation of Eqs. (\ref{sim1})-(\ref{sim3}) 
is presented in Appendix \ref{app S-S}.

We apply the zooming transformation only in
the $x$-direction but not in the $y$-direction.
This is because 
the gas contracts along $x$-axis but not along $y$-axis
in the unperturbed state.
In the ordinary coordinate, the transverse scale of the perturbation
is expected to be constant with time.
However, if the zooming transformation is also applied in the $y$-direction,
the transverse scale of the perturbation decreases with time
in the ordinary coordinate, although the unperturbed gas does not
contract along the $y$-axis. Therefore, we apply the zooming
transformation only in the $x$-direction.

\subsection{Review of self-similar solutions}
In the zooming coordinate, steady state solutions 
correspond to S-S solutions that were derived in IT08.
In this section, physical properties of the S-S solutions are 
reviewed briefly.

The S-S solutions are specified by two parameters, $\alpha$ and $\omega$.
For convenience, instead of $\omega$, we can use a parameter $\eta$, which 
is given by 
\begin{equation}
        \eta =  \frac{(2-\alpha)\{1-(3-2\alpha)\omega\}}{1-\omega}.
\end{equation}
Using these parameters ($\alpha,\eta$), 
the time dependences of the central density, $\rho_{00}$, and pressure,
$P_{00}$, are given by
\begin{equation}
\rho_{00}(t) \propto t_*^{-\alpha_{\rho}(\eta)}\;\;\;\mathrm{and}\;\;\;
        P_{00}(t)\propto t_*^{\alpha_P(\eta)},
\label{den00 P00}
\end{equation}
respectively, where $\alpha_\rho(\eta) = \eta/(2-\alpha)$ and $\alpha_P(\eta)
=(1-\eta)/(1-\alpha)$.

The S-S solutions include two asymptotic solutions.
For $\eta\sim0$, the time dependences of the central density and pressure are
given by
\begin{equation}
        \rho_{00}(t) \sim \mathrm{const.}\;\;\;\mathrm{and}\;\;\;
        P_{00}(t)\propto t_*^{1/(1-\alpha)},
\end{equation}
respectively.
This time evolution indicates the isochoric mode. 
For $\eta\sim1$, the time dependences of the central density and pressure 
are given by
\begin{equation}
        \rho_{00}(t) \propto t_*^{-1/(2-\alpha)}\;\;\;\mathrm{and}\;\;\;
        P_{00}(t)\sim\mathrm{const.},
\end{equation}
respectively.
This time evolution corresponds to the isobaric mode.
Our S-S solutions exist between the two limits, $0<\eta<1$.
With the condition that the 
increasing rate of $\rho_{00}(t)$ is equal to the decreasing rate of $P_{00}(t)$, or
$\alpha_\rho(\eta) = \alpha_P(\eta)$, 
the critical value of $\eta$ is given by
\begin{equation}
        \eta_\mathrm{eq} = (2-\alpha)/(3-2\alpha).
\end{equation}
Therefore, the S-S solutions for $0<\eta<\eta_\mathrm{eq}$ and $\eta_\mathrm{eq}<\eta<1$ are
close to the isochoric and the isobaric modes, respectively.
During S-S condensation, the central pressure should not increase with time.
From Eq. (\ref{den00 P00}),
the range of the value of $\alpha$ is found to be
$\alpha<1$.

\subsection{Perturbation equations}\label{per eq}
Perturbation on the S-S solutions is considered. 
Perturbed variables are defined by
\begin{eqnarray}
    \Omega &=& \Omega_0(\xi)\{1+\delta \Omega(\tau,\xi,y)\}, \nonumber\\
    V_x &=&  V_0(\xi) + \delta V_x(\tau,\xi,y), \nonumber\\
    V_y &=&  \delta V_y(\tau,\xi,y), \nonumber\\
    \mathrm{and} && \nonumber \\
    \Pi &=&  \Pi_0(\xi)\{1 + \delta \Pi(\tau,\xi,y)\},
    \label{per var}
\end{eqnarray}
where subscript "0" indicates the unperturbed state.
As the perturbation, we consider the following Fourier mode with respect to $y$,
\begin{equation}
        \vec{\delta Q}(\tau,\xi,y) = \vec{\delta Q}(\tau,\xi) \exp(iky),\;\;\;
        \mathrm{where\;\;}\vec{\delta Q} 
        = (\delta \Omega, \delta \vec{V},\delta \Pi),
        \label{kne0 Q}
\end{equation}
and $k$ indicates the wave number of the plane wave that propagates along $y$-direction.
Substituting Eqs. (\ref{per var}) and (\ref{kne0 Q}) into Eqs.
(\ref{sim1})-(\ref{sim3}) and linearizing, we get the following perturbation equations:
\begin{equation}
        \frac{\mathrm{D} \delta \Omega}{\mathrm{D} \tau}
        + \frac{\partial \delta V_x}{\partial \xi}
        = - (\ln \Omega_0)' \delta V_x - \kappa(\tau)i\delta V_y,
        \label{per eoc}
\end{equation}
\begin{equation}
        \frac{\mathrm{D} \delta V_x}{\mathrm{D} \tau}
        + \frac{X_0^2}{\gamma}\frac{\partial\delta \Pi}{\partial\xi}
        = -(\omega+V_0') \delta V_x- \frac{X_0^2}{\gamma}(\ln \Pi_0)'(
        \delta \Pi - \delta \Omega), 
        \label{per eomx}
\end{equation}
\begin{equation}
  \frac{\mathrm{D} i\delta V_y}{\mathrm{D} \tau}
        = -\omega i\delta V_y+ \kappa(\tau)\frac{X_0^2}{\gamma}\delta \Pi,
        \label{per eomy}
\end{equation}
and
\begin{eqnarray}
        &&  \hspace{-4mm} 
        \frac{\mathrm{D} \delta \Pi }{\mathrm{D} \tau}
        -\gamma
        \frac{\mathrm{D} \delta \Omega}{\mathrm{D} \tau}
        \nonumber\\
     && = -(2-\alpha)\gamma\epsilon_0\delta \Omega
         -(\alpha-1)\gamma\epsilon_0\delta \Pi 
         -(\ln \Pi_0\Omega_0^{-\gamma})'\delta V_x,
        \label{per eoe}
\end{eqnarray}
where $\mathrm{D}/\mathrm{D}\tau=\partial/\partial \tau+V_0
\partial /\partial \xi$, 
\begin{equation}
\epsilon_0 = \gamma^{\alpha-1}(\gamma-1)\Omega_0^{2-\alpha}\Pi_0^{\alpha-1},
\;\;\mathrm{and}\;\;\kappa(\tau) \equiv kx_0(\tau).
\end{equation}

The time-dependent factors remain in the form of $\kappa(\tau)=kx_0(\tau)$ 
in (\ref{per eoc})-(\ref{per eoe}),
because the transverse scale is not zoomed (see Eq. \ref{zooming}) 
as mentioned above.
The leads to a problem that
the perturbed variables cannot be expanded 
in the Fourier mode with respect to $\tau$ in general.

\begin{figure*}[t]
    \begin{center}
       \includegraphics{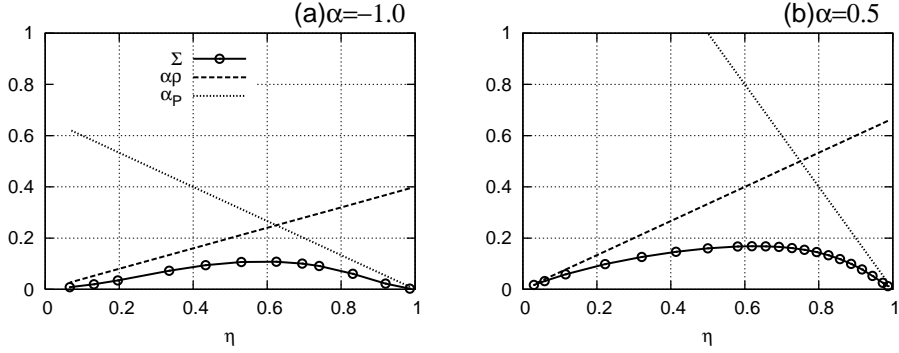}
        \caption{Growth rate, $\Sigma$, as a function
        of $\eta$
        for (a) $\alpha=-1.0$ and
        (b) 0.5 for the case with the perturbation parallel to the condensation.
        For comparison, the increasing rate of the unperturbed central density, 
        $\alpha_\rho(\eta)$, 
        and the decreasing rate of the unperturbed central pressure, 
        $\alpha_P(\eta)$, are shown by 
        the dashed and dotted lines, respectively.
        }
        \label{k0 grow}
    \end{center}
\end{figure*}
\section{Perturbation for $\vec{k=0}$}\label{keq0}

In this section, we consider the perturbation parallel to the condensation, or for 
the case with $k=0$.
In this case, since the time-dependent factor, $\kappa(\tau)=kx_0(\tau)$, vanishes, 
the perturbed variables can be expanded in the Fourier mode with respect to $\tau$ as
\begin{equation}
	\delta Q(\tau,\xi) = \delta Q(\xi) e^{\sigma \tau}.
  \label{fourier}
\end{equation}
By Eq. (\ref{fourier}), the time evolution of the perturbations
is given by 
\begin{equation}
        \delta Q\propto t_*^{-{\Sigma}},\;\;\;\mathrm{where}\;\;
        {\Sigma}=\frac{\sigma}{1-\omega}.
        \label{Sigma def}
\end{equation}
Substituting Eq. (\ref{fourier}) into the perturbation Eqs. (\ref{per eoc})-(\ref{per eoe}), 
one obtains the following ordinary differential equations:
\begin{equation}
      \frac{\mathrm{d}\delta Q_i}{\mathrm{d}\xi} = 
      \frac{1}{V_0^2-X_0^2}\sum_{j=1}^3 A_{ij}\delta Q_j,\;\; 
      \vec{\delta Q} = (\delta \Omega, \delta V_x, \delta \Pi),
     \label{keq0 eq}
\end{equation}
where the detailed expression of $A_{ij}$ is shown in Appendix \ref{app per}.
Equations (\ref{keq0 eq}) are solved as a boundary- and eigenvalue problem.

\subsection{Boundary conditions}\label{keq0 boundary}
We impose the boundary conditions at $\xi=0$ and at the critical point, $\xi=\xi_\mathrm{s}$, where
$V_0=X_0$. 
The boundary conditions at $\xi=0$ are obtained by the asymptotic limit of the perturbed variables.
From the regularity of the perturbed variables at $\xi=0$, we find that 
perturbed variables should have the following asymptotic forms:
\begin{equation}
        \lim_{\xi\rightarrow0}\delta \Omega(\xi) \simeq \delta \Omega_0,
        \;\;\;\lim_{\xi\rightarrow0}\delta V_x(\xi) \simeq -\sigma\delta \Omega_0 \xi,
     \label{keq0 den}
\end{equation}
and
\begin{equation}
        \lim_{\xi\rightarrow0}
        \delta \Pi(\xi) \simeq \gamma\frac{\sigma + (\alpha-2)\epsilon_{00}}
        {\sigma + (\alpha-1)\gamma\epsilon_{00}}\delta \Omega_0,
     \label{keq0 p}
\end{equation}
where $\epsilon_{00}=\{2\omega - \beta(\gamma -1)\}/\gamma$

The boundary conditions at the critical point, $\xi=\xi_\mathrm{s}$, are obtained 
from the Eqs. (\ref{keq0 eq}).
At $\xi=\xi_\mathrm{s}$, the denominator of the righthand side becomes zero.
To obtain a regular solution from $\xi=0$ to $\xi=\infty$, 
the numerator of the righthand side should vanish.
Therefore, the boundary conditions are given by the following three equations,
\begin{equation}
        \sum_{j=1}^3 A_{ij}\delta Q_j\Biggr|_{\xi=\xi_\mathrm{s}}=0,\;\; i=1,2,3.
\label{bound cri k=0}
\end{equation}
The above three equations give only one independent condition.

\subsection{Numerical method}\label{num}
Solutions of Eqs. (\ref{keq0 eq}) have three 
integration constants. Therefore, if we set two constants
$(\delta \Omega_0,\sigma)$ and impose the boundary condition at $\xi=\xi_\mathrm{s}$, the solution 
is completely fixed. In this section, our numerical method 
for solving Eqs. (\ref{keq0 eq}) is described.

We can set $\delta \Omega_0=1$ without loss of generality. 
For a given $\sigma$, we integrate Eqs. (\ref{keq0 eq}) from $\xi=0$ to 
the critical point, $\xi=\xi_\mathrm{s}$, using a fourth-order Runge-Kutta method.
Eigenvalue, $\sigma$, is modified until the perturbed variables satisfy 
the boundary condition at $\xi=\xi_\mathrm{s}$
using the Newton-Raphson method.
After that, we integrate Eqs. ($\ref{keq0 eq}$) up to $\xi=10^4$.

\subsection{Results}
Figure \ref{k0 grow} 
shows the dependence of growth rate, $\Sigma$, on
$\eta$ for $(a)\alpha$ = -1.0 and (b)0.5.
From Fig. \ref{k0 grow}, 
it is seen that $\sigma>0$ for a wide range of $\alpha$ and $\eta$. 
Therefore, the perturbation is unstable. 
However, $\sigma$ is smaller than $\alpha_P$ and $\alpha_\rho$.
Therefore, the growth rate is too low to grow sufficiently during runaway condensation.
This is consistent with the results of the one-dimensional numerical simulation 
shown in Sect. \ref{astro impli}.

%
\section{Perturbation with $\vec{k\ne 0}$}\label{sec kne0}
For $k\ne 0$, the perturbation equations contain the time-dependent factor,
$\kappa(t)$.
Here we show that this factor, $\kappa=kx_0(t)$, is related to the ratio of 
the local cooling length at the centre to 
the wave length of the perturbation. 
The cooling timescale at $t=0$ is given by
\begin{equation}
        t_\mathrm{cool}^{0} 
        = \frac{P_{00}}{(\gamma-1)\gamma^\alpha\Lambda_0 \rho_{00}^{2-\alpha}P_{00}^\alpha}
        = \frac{1-\omega}{\gamma^\alpha(\gamma-1)}
        \Omega_{00}^{\alpha-2}\Pi_{00}^{1-\alpha}t_\mathrm{c},
\end{equation}
where superscript "0" indicates the value at $t=0$.
Therefore, using $t_\mathrm{cool}^{0}$, the collapse time can be expressed as 
\begin{equation}
        t_\mathrm{c} = 
        \frac{\gamma^\alpha(\gamma-1)}{1-\omega}\Omega_{00}^{2-\alpha}\Pi_{00}^{\alpha-1}
        t_\mathrm{cool}^{0}.
 \label{tc}
\end{equation}
Using Eqs. (\ref{per vari}) and (\ref{self vari1}), at $t=0$, 
the sound speed at the centre is given by 
\begin{equation}
        c_{00}^0 = X_{00}v_0^0,\;\;\mathrm{where}\;\;v_0^0t_\mathrm{c} =
        a/(1-\omega) = x_0(0)/(1-\omega).
\end{equation}
Therefore, $x_0(0)$ is given by
\begin{equation}
        x_0(0) = \frac{1-\omega}{X_{00}}c_{00}^0 t_\mathrm{c}.
\label{x0}
\end{equation}
Using Eq. (\ref{tc}), Eq. (\ref{x0}) can be written as 
\begin{equation}
  x_0(0)  = M_{00} L_\mathrm{cool}^0=a,
\end{equation}
where $M_{00}= \gamma^{\alpha-1/2}(\gamma-1)
\Omega_{00}^{5/2-\alpha}\Pi_{00}^{\alpha-3/2}$, and 
$L_\mathrm{cool}^{0} = c_{00}^{0} t_\mathrm{cool}^{0}$
is the cooling length at the centre at $t=0$.
Since the S-S solutions are invariant under the transformation 
$\xi\rightarrow m\xi$, $V_0\rightarrow mV_0$, 
$\Omega_0\rightarrow m^{-2(\alpha-1)}
\Omega_0$, and $\Pi_0\rightarrow m^{-2(\alpha-2)}\Pi_0$, with
a parameter $m$, one can take $m$ in such a way that 
$M_{00}$ is equal to 1.
Hereafter, $M_{00}$ is set to unity.
Therefore, the time-dependent factor, $\kappa(\tau)$, is expressed as 
\begin{equation}
 \kappa(\tau) = k x_0(\tau) = k L_\mathrm{cool}^{0} t_*^{1/(1-\omega)} =
        kL_\mathrm{cool}^0e^\mathrm{-\tau}
         = k L_\mathrm{cool}(\tau),
\label{kappa}
\end{equation}
where $L_\mathrm{cool}(\tau)=L_\mathrm{cool}^{0}e^{-\tau}$.

\subsection{Static approximation}\label{kneq0 eigen}
\begin{figure*}[t]
    \begin{center}
       \includegraphics{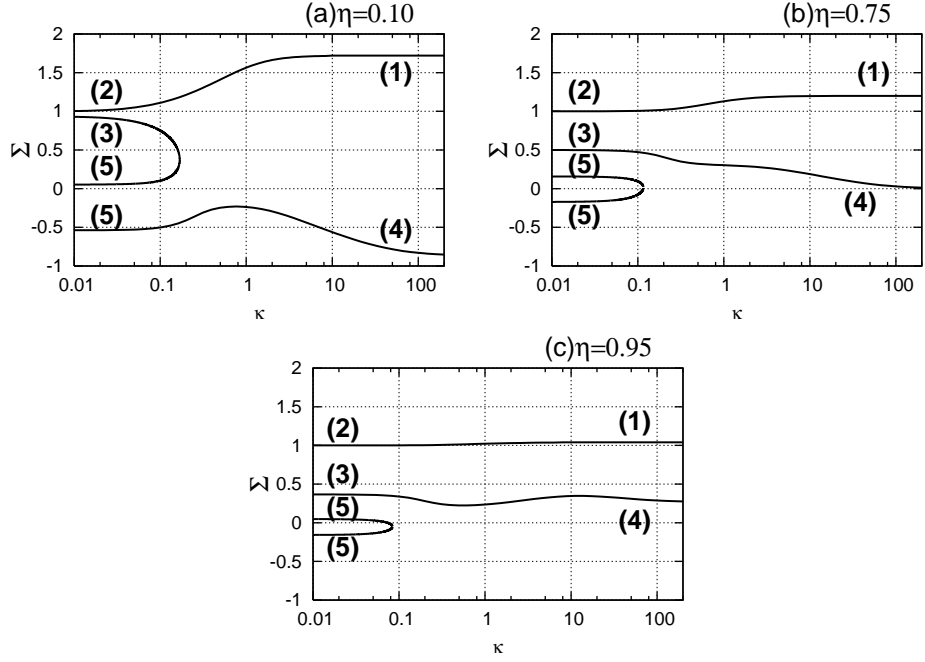}
        \caption{Approximate dispersion relations for $\alpha=0.5$ with
        (a) $\eta=0.1$, (b) 0.75, and (c) 0.95, where $\kappa$ is 
        the nondimensional wave number, and $\Sigma$ 
        the growth rate of the perturbation.
        The labels of number represent the branches of modes.
        }
        \label{kneq0 disp}
    \end{center}
\end{figure*}

Before we present the fully time-dependent numerical calculation 
in Sect. \ref{numerical cal}, 
we at first consider a special case in which the time evolutions of the unperturbed
S-S solutions are slower than the growth of perturbation.
In this situation, the time evolution of the unperturbed state
is negligible during the growth of the perturbations.
Therefore, we set $x_0$ to be a constant in Eqs.
(\ref{per eoc})-(\ref{per eoe}).
This approximation is also valid for $k\ll1/x_0$ where the term, $\kappa$, is negligibly small
in Eqs. (\ref{per eoc})-(\ref{per eoe}).
We use the Fourier mode as 
\begin{equation}
        \vec{\delta Q}(\xi,y,\tau) = \vec{\delta Q}(\xi) e^{iky + \sigma \tau}
       ,\;\;\;\vec{\delta Q} = (\delta \Omega, \delta V_x,\delta\Pi,i\delta V_y).
  \label{fourier kne0}
\end{equation}
The condition under which the static approximation is valid is given by 
\begin{eqnarray}
        \left| \frac{\mathrm{d}\ln \delta Q}{\mathrm{d}\ln t_*} \right| = 
        \Sigma  \gg
        \left| \frac{\mathrm{d}\ln \kappa}{\mathrm{d}\ln t_*} \right|
       & =&\frac{(2-\alpha)(3-2\alpha) - \eta}{2(2-\alpha)(1-\alpha)},\nonumber\\
     &\hspace{-2cm}  =& \hspace{-1cm}  \left\{\begin{array}{cc}
               (5-2\alpha)/\left\{ 2(2-\alpha) \right\}& \;\;\mathrm{for}\;\;\eta=1 \\
               (3-2\alpha)/\left\{ 2(1-\alpha) \right\}& \;\;\mathrm{for}\;\;\eta=0 \\
        \end{array}\right.,
        \label{cond}
\end{eqnarray}
where the definition of $\Sigma$ is the same as that in Eq. (\ref{Sigma def}).
Substituting Eq. (\ref{fourier kne0})
into the perturbation Eqs. (\ref{per eoc})-(\ref{per eoe}), 
we get 
\begin{equation}
        \frac{\mathrm{d}\delta Q_i}{\mathrm{d}\xi} = 
        \frac{1}{V_0^2 - X_0^2 } \sum_{j=1}^4 A_{ij}\delta Q_j
       ,\;\;\;\vec{\delta Q} = (\delta \Omega, \delta V_x,\delta\Pi,i\delta V_y),
\label{kne0 eq}
\end{equation}
where the detailed expression of $A_{ij}$ is shown in Appendix \ref{app per}.

We impose the boundary conditions at $\xi=0$ and $\xi=\xi_\mathrm{s}$.
Since we are interested in the fragmentation of the cooling layer, 
only the even mode is investigated.
For the even mode,
the perturbed variable should have the following asymptotic forms in $\xi\ll1$: 
\begin{eqnarray}
        \lim_{\xi\rightarrow0}\delta \Omega(\xi) & \simeq&  \delta \Omega_0, \nonumber\\
        \lim_{\xi\rightarrow0}\delta V_x(\xi) & \simeq&  \delta V_{x0}\xi, \nonumber\\
        \lim_{\xi\rightarrow0}\delta V_y(\xi) &\simeq &  \delta V_{y0}, \nonumber\\
        && \hspace{-2cm}\mathrm{and} \nonumber \\
        \lim_{\xi\rightarrow0}\delta \Pi(\xi) &\simeq &  \delta \Pi_0 + 
        \delta \Pi_{01}\xi^2.
        \label{asy kneq0}
\end{eqnarray}
Substituting Eqs. (\ref{asy kneq0}) into Eqs. (\ref{kne0 eq}), we obtain 
the following relations:
\begin{equation}
        \sigma\Omega_0 + \delta V_{x0} + \kappa i\delta V_{y0}=0,
        \label{kneq0 app eoc}
\end{equation}
\begin{equation}
        (\sigma-\omega)i\delta V_{y0} = 
        \frac{\kappa X_{00}^2}{\gamma}\delta \Pi_0,
        \label{kneq0 app vy}
\end{equation}
\begin{equation}
        \left\{\sigma + (\alpha-2)\epsilon_{00}\right\}\gamma\delta \Omega_0
        - \left\{ \sigma + (\alpha-1)\gamma\epsilon_{00} \right\}\delta \Pi_0=0,
        \label{kneq0 app eoe}
\end{equation}
and 
\begin{equation}
        (2\beta+1 - \omega + \sigma )\delta V_{x0} + 2\frac{X_{00}^2}{\gamma}
        \delta \Pi_{01} = 2\beta\omega(1-\alpha)(\delta \Pi_0 - \delta \Omega_0).
        \label{kneq0 app eom}
\end{equation}
The boundary condition at $\xi=\xi_\mathrm{s}$ is derived in the same way
in Sect. \ref{keq0 boundary}, and we obtain two independent conditions. 
Numerical method for solving Eqs. (\ref{kne0 eq}) is the same as that in
Sect. \ref{num}.

Figures \ref{kneq0 disp} shows an approximate dispersion relation
for $\alpha=0.5$ with (a)$\eta=0.95$, (b) 0.75, and (c) 0.10.
In Fig. \ref{kneq0 disp}, one can see several branches labelled by numbers.
The most unstable branch is labelled by (1) for large $\kappa$ limit, and (2) for small $\kappa$
limit.
Each branch is explained below.\\

\begin{figure}[h]
        \begin{center}
                \includegraphics[width=6cm,height=4cm]{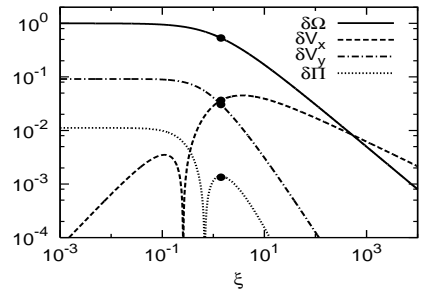}
        \end{center}
        \caption{Eigenfunctions for $\eta=0.75$ and $\kappa=8$. 
        The filled circles indicate the values at the critical point. 
        The eigenvalue, $\Sigma$, is equal to 1.196.}
        \label{isobaric eigen func}
\end{figure}

{\bf (1) The isobaric mode}
\begin{figure*}[t]
        \begin{center}
                \includegraphics{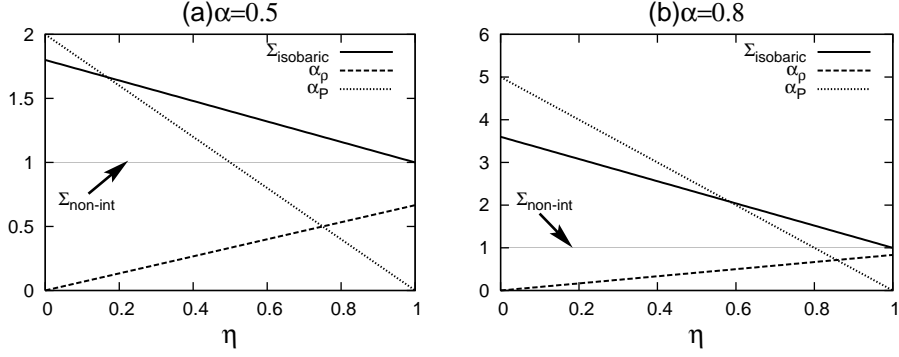}
        \end{center}
        \caption{Growth rate as a function of 
        $\eta$ for (a)$\alpha$ = 0.5 and (b)0.8 for the case with $k\ne0$.
        The thick solid lines correspond to the growth rate of the isobaric mode, $\Sigma_\mathrm{isobaric}$. 
        For comparison, the evolutionary rate 
        of the unperturbed central density, $\alpha_\rho(\eta)$,
        and pressure, $\alpha_P(\eta)$ are shown by the dashed and dotted lines, respectively.
        The thin solid lines pointed by arrows correspond to 
        the growth rate in the noninteractive mode, $\Sigma_\mathrm{non-int}$.
        }
        \label{sig isob}
\end{figure*}\\
The branch (1) corresponds to the most unstable mode for 
$\kappa\gg1$, or $1/k\ll L_\mathrm{cool}$.
Since $1/k\ll L_\mathrm{cool}$, 
the sound wave can travel the wave length of the perturbation many times
during the runaway cooling of the unperturbed state.
Therefore, the perturbation is expected to grow in pressure 
equilibrium with its surroundings, 
and the mode corresponds to the isobaric mode.
An example of eigenfunctions $(\delta \Omega, \delta V_x, 
\delta V_y, \delta \Pi)$ for the branch (1) 
is shown in Fig. \ref{isobaric eigen func} for $\eta=0.75$ and $\kappa=8.0$.
In Fig. \ref{isobaric eigen func}, it is clearly seen that 
$|\delta \Pi|\ll|\delta \Omega|$.
This behaviour also implies the isobaric mode.

The growth rate in the isobaric mode can also be derived analytically 
by considering the evolution of a fluid element at the centre.  The fluid element 
is assumed to have an isobaric fluctuation, 
$\rho=\rho_{00}(t) + \delta \rho_{00}$ and $P=P_{00}(t)$. 
From Eq. (\ref{eoe}), 
the following perturbation equation is obtained:
\begin{equation}
        \frac{\partial}{\partial t}\left( \frac{\delta \rho_{00}}{\rho_{00}} \right)
        = (2-\alpha)\gamma^{\alpha-1}(\gamma-1)\Lambda_0 \rho_{00}^{2-\alpha}P_{00}^{\alpha-1} 
        \frac{\delta\rho_{00}}{\rho_{00}}.
        \label{isobaric grow}
\end{equation}
From Eqs. (\ref{self vari2}) and (\ref{self vari3}), we have
\begin{equation}
        \gamma^{\alpha-1}(\gamma-1)\Lambda_0 \rho_{00}^{2-\alpha}P_{00}^{\alpha-1} 
        = \frac{\epsilon_{00}}{(1-\omega)(t_\mathrm{c}-t)}.
        \label{rel}
\end{equation}
Using Eq. (\ref{rel}), Eq. (\ref{isobaric grow}) is rewritten as
\begin{equation}
        \frac{\partial}{\partial t}\left( \frac{\delta \rho_{00}}{\rho_{00}} \right)
        = \frac{(2-\alpha)\epsilon_{00}}{(1-\omega)(t_\mathrm{c}-t)}
        \frac{\delta\rho_{00}}{\rho_{00}}.
        \label{isobaric grow1}
\end{equation}
Equation (\ref{isobaric grow1}) can be integrated to give
\begin{equation}
        \frac{\delta \rho_{00}}{\rho_{00}}\propto t_*^
        {-(2-\alpha)\epsilon_{00}/(1-\omega)}.
\end{equation}
Therefore, the growth rate in isobaric mode is given by 
\begin{eqnarray}
        \Sigma_\mathrm{isobaric} = \frac{(2-\alpha)\epsilon_{00}}{1-\omega} && 
        \nonumber \\
        &&
       \hspace{-28mm} =\left[ 1-\frac{2-\alpha}{\gamma(1-\alpha)} \right]\eta
        + \frac{2-\alpha}{\gamma(1-\alpha)}
        =\left\{
        \begin{array}{cc}
                1.72 & \;\;\mathrm{for}\;\;\eta=0.10 \\
                1.20 & \;\;\mathrm{for}\;\;\eta=0.75 \\
                1.04 & \;\;\mathrm{for}\;\;\eta=0.95 \\
        \end{array}
        \right..
 \label{isob growth}
\end{eqnarray}
In Fig. \ref{kneq0 disp}b, the eigen-value, $\Sigma$, is found to be 1.196 
for the case with $\eta=0.75$ and $\kappa=8.0$.
This is very close to 
$\Sigma_\mathrm{isobaric}=1.2$ for $\eta=0.75$.
From Fig. \ref{kneq0 disp}, it is clearly seen that the growth rate approaches
the corresponding $\Sigma_\mathrm{isobaric}$ in the large $\kappa$ limit.
The analytic growth rate is derived by the local argument.
Therefore, the growth rate is expected to be
independent of a global structure of the system.
\citet{BL00} performed a linear analysis on a spatially uniform and 
isochorically cooling gas. Their growth rate in the isobaric mode 
and our $\Sigma_\mathrm{isobaric}$ 
for $\eta=0$ are identical, although the spatial structure
of the unperturbed state is quite different.

For the density perturbation to grow sufficiently during the runaway cooling,
it must grow faster than the condensation of the cooling layer.
This condition can be expressed by $\Sigma_\mathrm{isobaric}>\alpha_\rho$.
Figure \ref{sig isob} shows $\Sigma_\mathrm{isobaric}$, 
$\alpha_\rho$ and $\alpha_P$ as a function of $\eta$ 
for (a) $\alpha$=0.5 and (b) 0.8.
From Fig. \ref{sig isob}, it is found that $\Sigma_\mathrm{isobaric}$ is greater than $\alpha_\rho$ 
for all $\eta$.
For other $\alpha$, we can investigate analytically as follows:
subtracting $\alpha_\rho$ from $\Sigma_\mathrm{isobaric}$, one obtains
\begin{eqnarray}
        \Sigma_\mathrm{isobaric} - \alpha_\rho &=& 
        \left[ 1-\frac{2-\alpha}{\gamma(1-\alpha)}  - \frac{1}{2-\alpha}\right]\eta
        + \frac{2-\alpha}{\gamma(1-\alpha)} \nonumber\\
        &\ge&1-\frac{1}{2-\alpha}>0\;\;\mathrm{for\;\;\alpha<1}.
\end{eqnarray} 
Therefore, for $\alpha<1$, $\Sigma_\mathrm{isobaric}$ is more than $\alpha_\rho$, and 
the isobaric mode can grow in the runaway cooling layer. \\

\begin{figure}[h]
        \begin{center}
                \includegraphics[width=8cm]{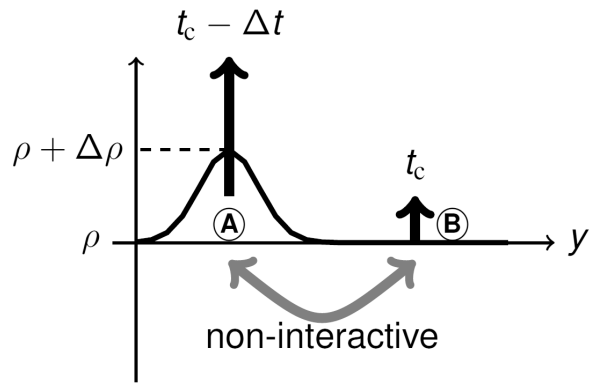}
        \end{center}
        \caption{Schematic picture of the noninteractive mode. }
        \label{schematic}
\end{figure}

{\bf (2) The noninteractive mode}

Branch (2) corresponds to the most unstable mode for $\kappa\ll1$.
Because the wave length is larger than the cooling length, each part evolves independently 
according to the S-S solutions. We call this branch the noninteractive mode.
Figure \ref{schematic} shows the schematic picture of the noninteractive mode.
In Fig. \ref{schematic}, similarity variables have an initial fluctuation.
For example, we consider two different regions, "A" and "B", where 
$\rho_\mathrm{A}=\rho +\Delta \rho$ and 
$\rho_\mathrm{B}=\rho$.
Due to the difference of $\Delta \rho$, 
the regions "A" and "B" have different collapse time,
$t_\mathrm{c}-\Delta t$ and $t_\mathrm{c}$, respectively.
Omitting any terms that do not grow, we find the time evolution of difference, $\Delta \rho$, to be
\begin{equation}
\frac{\Delta \rho}{\rho_0(t)}  =
- \Omega(\xi)\Delta t\left[ \alpha_\rho + 
\frac{1}{1-\omega}\frac{\mathrm{d} \ln\Omega}{\mathrm{d}\ln\xi} \right]
\frac{1}{t_\mathrm{c}-t}.
\end{equation}
Therefore, in the zooming coordinate, 
the density perturbation grows as $\delta \Omega(\xi)\propto
(t_\mathrm{c}-t)^{-1}$. Other perturbed variables also grow in the same power law.
Therefore, comparing with Eq. (\ref{Sigma def}), the growth rate is given by 
\begin{equation}
        \Sigma_\mathrm{non-int} =1,
\end{equation}
which is independent of $\alpha$ and $\eta$.
The noninteractive mode arises from the fluctuation of the collapse time, $t_\mathrm{c}$, 
due to density and pressure perturbations.
From the physical mechanism, 
the perturbation grows in the same way as the unperturbed state.
In other words, the perturbation of the isobaric (isochoric) 
cooling layer grows isobarically (isochorically).

We investigate whether the noninteractive mode grows sufficiently during the runaway cooling or not.
First, we consider the case with $\eta> \eta_\mathrm{eq}$.
Since the perturbation grows isobarically, $\Sigma_\mathrm{non-int}$ is compared to
$\alpha_\rho=\eta/(2-\alpha)$. 
From Fig. \ref{sig isob}, it is seen that the growth rate, $\Sigma_\mathrm{non-int}$, is 
higher than $\alpha_\rho$ for all $\eta>\eta_\mathrm{eq}=$ 0.75 and 0.86 
for $\alpha=0.5$ and 0.8, respectively.
Therefore, the noninteractive mode can grow sufficiently.
Next, we consider the cooling layer with $\eta<\eta_\mathrm{eq}$. 
In this layer, since the perturbation grows
isochorically, $\Sigma_\mathrm{non-int}$ is compared to $\alpha_P=(1-\eta)/(1-\alpha)$.
Analytically, it is found that the pressure perturbation 
can only grow for $\eta>\alpha$. 

\cite{KI00} performed
a linear analysis of a spatially uniform and isobarically cooling gas in their appendix.
However, they did not find the noninteractive mode. 
This is because they fixed 
the collapse time to be 
spatially constant, and it was assumed not to be influenced by perturbation. 
\cite{BL00} performed
linear analysis of a spatially uniform and isochorically cooling gas by taking account
of the time evolution of the unperturbed state.
They showed that a perturbation cannot grow in the condition for long wave length limit.
Our result is consistent with theirs.
\\

{\bf (3) The shear mode}

For $\kappa\ll0$, there is a solution where physical variables are very small except for $\delta V_y$ 
which is spatially constant, and the eigenvalue is $\sigma=\omega$.
The similar mode was found by \cite{M93}, who investigated TI of a uniform granular medium. 
\cite{M93} called this mode the shear mode.
The growth rate is given by 
\begin{equation}
        \Sigma_\mathrm{shear}=\frac{\omega}{1-\omega}
        = \left\{
        \begin{array}{cc}
               0.93 & \;\;\mathrm{for}\;\;\eta=0.10 \\
               0.50 & \;\;\mathrm{for}\;\;\eta=0.75 \\
               0.37 & \;\;\mathrm{for}\;\;\eta=0.95 \\
        \end{array}
       \right..
        \label{growth rate shear}
\end{equation}
In Fig. \ref{kneq0 disp}, it is seen that each growth rate in branch (3) has the corresponding value of 
$\Sigma_\mathrm{shear}$ for $\kappa\ll1$.
The physical meaning of this mode can be understood as follows:
for $\kappa\ll 1$, since the effect of the pressure gradient with respect to $y$ is very 
weak, the gas can freely stream with almost constant velocity, $v_y$, in the $y$ direction.
On the other hand, the central sound speed, $c_{00}(t)$, decreases 
as $\propto (t_\mathrm{c} - t)^{\omega/(1-\omega)}$.
Therefore, the ratio of the dynamical velocity to the thermal velocity, $v_y/c_{00}(t)$,
grows with time as $\propto (t_\mathrm{c} - t)^{-\omega/(1-\omega)}$, indicating that the 
growth rate is given by Eq. (\ref{growth rate shear}).
For the case with larger wave number, the effect of the pressure gradient becomes important.
Therefore, the fluid element cannot stream freely, and the growth rate is lower
as shown in Fig. \ref{kneq0 disp}.
\\

{\bf (4) The free-streaming mode}

For large $\kappa$, there is another mode in which 
the velocity perturbation in the $x$-direction is 
much greater than that in the $y$-direction, $|\delta V_{x0}|\gg |\delta V_{y0}|$.
We call this mode the free-streaming mode.
From Eq. (\ref{kneq0 app eom}) with $\delta \Omega_0=\delta \Pi_0=\delta \Pi_{01}=0$, we obtain 
\begin{equation}
        \Sigma_\mathrm{free} = -1 - \frac{2\beta}{1-\omega}
        = \left\{
        \begin{array}{cc}
              -0.87 & \;\;\mathrm{for}\;\;\eta=0.10 \\
               0.00 & \;\;\mathrm{for}\;\;\eta=0.75 \\
               0.27 & \;\;\mathrm{for}\;\;\eta=0.95 \\
        \end{array}
        \right..
\end{equation}
In the free-streaming mode, the growth of the velocity perturbation in the $x$-direction 
is hampered by the pressure gradient of the unperturbed state.
Therefore, the growth rate is less than the shear mode.\\

{\bf (5) $\vec{k=0}$ mode}\\
The growth rate in this branch for $\kappa\ll1$ 
coincides with the case with $k=0$, which is obtained in 
Sect. \ref{keq0}.\\

\begin{figure*}[t]
    \begin{center}
      \begin{tabular}{cc}
       \includegraphics{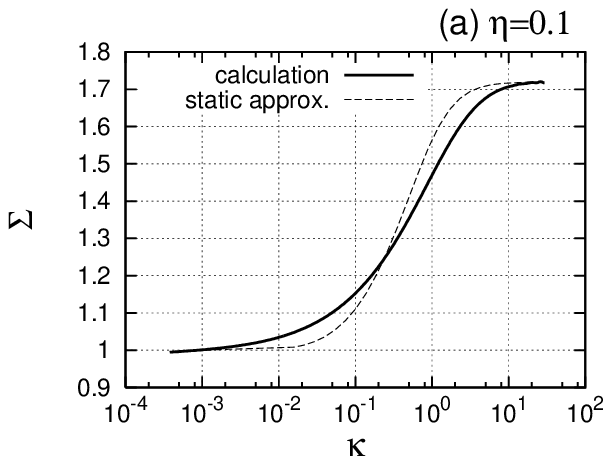}
       &
       \includegraphics{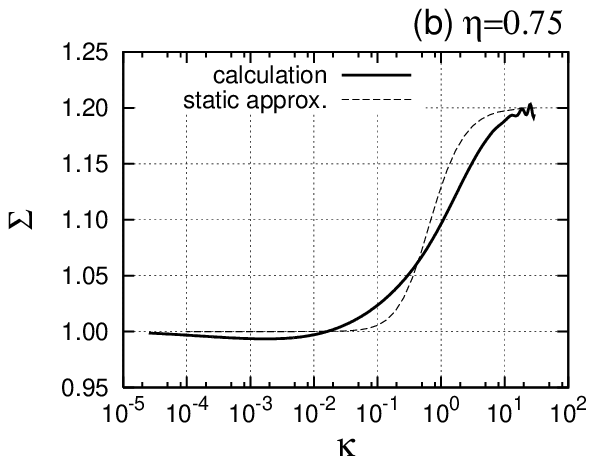}\\
      \end{tabular}
        \caption{Growth rate obtained by results of 
        numerical linear analysis for $\eta$= 0.1(a)  and 0.75(b). 
        The thick and dashed lines indicate the results
        of numerical linear analysis and those of approximate linear analysis.}
        \label{kneq0 grow}
    \end{center}
\end{figure*}
\subsection{Linear analysis considering the time evolution of $\kappa$}
\label{numerical cal}
The static approximation is valid only if $\Sigma$ is much larger than $|\mathrm{d}
\ln\kappa/\mathrm{d}\ln t_*|$.
However, 
from Fig. \ref{kneq0 disp}, it is found that the growth rate of the most unstable mode 
for each $\kappa$ is smaller than $|\mathrm{d} \ln\kappa/\mathrm{d}\ln t_*|$ whose
values are 1.93, 1.5, and 1.37 for 
$\eta=0.1$, 0.75, and 0.95 with $\alpha=0.5$, respectively [see Eq. (\ref{cond})].
Therefore, in this section, we
perform a linear analysis considering 
the time evolution of $\kappa(\tau)$ using direct numerical integration.
The upwind difference method is used as the numerical method to solve 
the perturbation equation.
The region of calculation is from $\xi=0$ to $\xi=100$ in the zooming coordinate.

We impose the boundary conditions at $\xi=0$ and $\xi=100$.
The even mode is set as the boundary condition at $\xi=0$.
The free boundary condition is set at $\xi=100$, but it does not influence the inner region
since the gas flows out supersonically 
from the outer boundary of the zooming coordinate.
As an initial state, we adopt the eigenfunction
of the isobaric mode for $\kappa=30$ which 
is obtained in Sect. \ref{kneq0 eigen}.

By solving Eqs. (\ref{per eoc})-(\ref{per eoe}), 
a time evolution of $\vec{\delta Q}$ is obtained.
During the calculation, the growth rate at $\tau$ is evaluated by 
\begin{equation}
        \Sigma_\mathrm{num} = \frac{1}{1-\omega}\frac{\mathrm{d} }{\mathrm{d}\tau}
        \left\{\ln \delta \Omega(\xi=0,\tau)\right\}
\end{equation}
at each instant of time.
For comparison with the result of the static approximation,
we focus on a relation between $\kappa(t)$ and the growth rate of 
the density perturbation at the centre, $\Sigma_\mathrm{num}$.
Figure \ref{kneq0 grow} shows the growth rate $\Sigma_\mathrm{num}$
as a function of $\kappa(\tau)$ at each instant of 
time for $\alpha=0.5$ with (a)$\eta$=0.1 and (b)0.75.
The nondimensional wave number, $\kappa(\tau)$, 
decreases with time as shown in Eq. (\ref{kappa}).
Therefore, in Fig. \ref{kneq0 grow}, 
the direction of time is from the right to the left.
For comparison, the approximate dispersion relations of branches (1)-(2)
in Fig. \ref{kneq0 disp} are superimposed by the dashed lines.
In both of Figs. \ref{kneq0 grow}a and \ref{kneq0 grow}b, the behaviour of the growth rate,
$\Sigma_\mathrm{num}$, moderately agrees with that of the approximated dispersion relations.
For $\kappa\gg1$, or initial phase, the growth rate agrees with $\Sigma_\mathrm{isobaric}$.
This is because the growth rate does not depend on $\kappa$ in the short wave length limit.
As $\kappa$ decreases and reaches about 1, $\Sigma_\mathrm{num}$ begins to decrease.
For $\kappa\ll1$, 
$\Sigma_\mathrm{num}$ approaches asymptotically 
$\Sigma_\mathrm{non-int}$, where the effect of $\kappa$ is negligible.
The effect of time-depending $\kappa$ is notable only for $0.1<\kappa<10$. Smoother dependence of 
the growth rate on $\kappa$ is obtained than the approximate dispersion relation.

\section{Discussion}\label{discuss}
\subsection{Astrophysical implication}\label{astro impli}
In this section, we estimate the fragmentation time of 
the cooling layer formed by a collision of WNM.
We consider a head-on collision between two thermally equilibrium gases
in which density and pressure 
are $\rho_\mathrm{WNM}=0.57m_\mathrm{H}\;\mathrm{cm}^{-3}$ and 
$P_\mathrm{WNM}=3.5\times10^3k_\mathrm{B}\;\mathrm{dyne\;cm^{-2}}$.
Two clouds collide along the $x$-axis at $t=0$ and $x=0$ with 
velocity $20\;\mathrm{km\;s^{-1}}$, i.e.,  
the mach number is 2.17.
The cooling function in this temperature region 
fitted by \citet{KI02} as follows:
\begin{equation}
        \rho{\cal L} = n (-\Gamma + n\Lambda)\;\;\mathrm{erg\;cm^{-3}\;s^{-1}},
\end{equation}
where
\begin{equation}
        \Gamma = 2\times10^{-26},
\end{equation}
\begin{equation}
        \frac{\Lambda}{\Gamma} = 
        1.0\times10^7 \exp\left( -\frac{118400}{T+1000} \right) + 
        1.4\times10^{-2}\sqrt{T}\exp\left( -\frac{92}{T} \right),
\end{equation}
where $n$ is the number density of the gas.
We perform numerical calculation using the one-dimensional Lagrangian Godunov 
scheme \citep{L97} between $x=0$ and $x=L_x = 3.26$pc.
At $x=0$ and $x=L_x$, we adopt the reflecting and free boundary conditions, respectively.
As the initial condition, we add the following density fluctuation:
\begin{equation}
        \rho(t=0,x)=\frac{\rho_\mathrm{WNM}}
        {\displaystyle 1+\frac{A}{i_\mathrm{max}}\sum_{i=0}^{i_\mathrm{max}} 
        \sin\left( \frac{2\pi }{L_x} 
        +\theta_i\right)    },
\end{equation}
where we set $i_\mathrm{max}=10,\;A=0.5$, and 
the phase, $\theta_i$, is given by random number between 0 and $2\pi$.
The maximum and minimum number density at the initial state are 
0.72 cm$^{-3}$ and 0.44 cm$^{-3}$, respectively.

\begin{figure}[h]
        \begin{center}
                \includegraphics{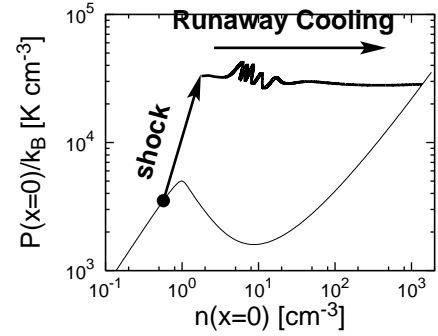}
        \end{center}
        \caption{Evolutionary track of the gas at 
        the centre after the head-on collision 
        (the thick solid line). The 
        ordinate and abscissa axes indicate the pressure and the
        number density,
        respectively. 
        The thin solid line indicates the thermally equilibrium curve.
        The filled circle represents the gas at the preshock region.
        }
        \label{Pden evo}
\end{figure}

\begin{figure}[h]
        \begin{center}
                \includegraphics{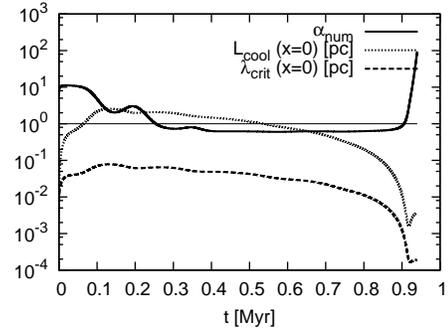}
        \end{center}
        \caption{
        Time evolution of $\alpha_\mathrm{num}$ (the solid line) 
        and the cooling length (the dotted line), which are evaluated at the centre .
        The dashed line indicates the time evolution of 
        the critical wave length defined in Sect. \ref{thermal conduction}.
        }
        \label{alpha evo}
\end{figure}

\begin{figure*}[t]
        \begin{center}
                \includegraphics{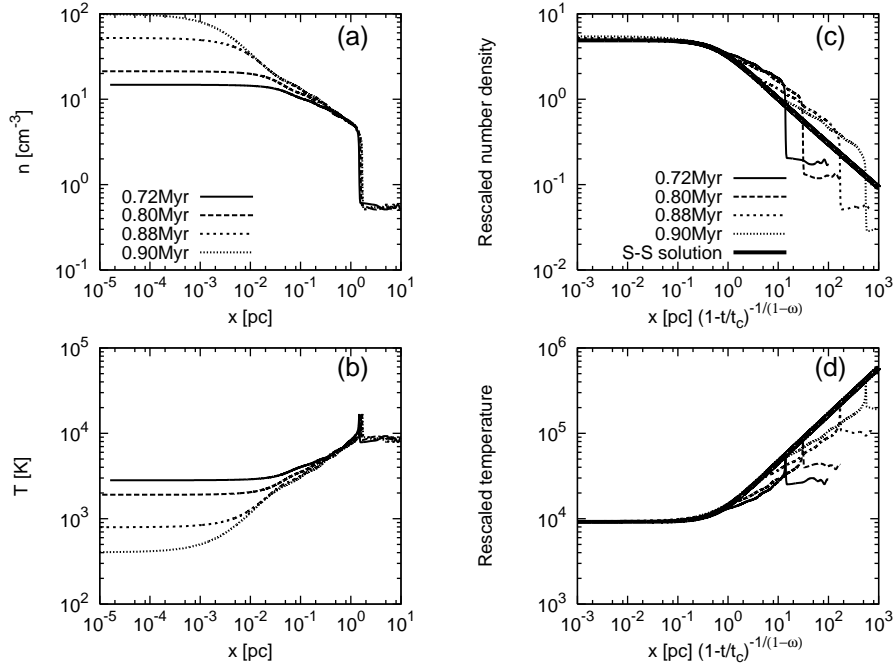}
        \end{center}
        \caption{
        Time evolution of (a)the number density and (b)the temperature, 
        respectively.
        The rescaled (c) number density, $nt_*^{\eta/(2-\alpha)}$ 
        and (d) temperature, $T t_*
        ^{(1-\alpha)(1-\eta)/\{\eta(2-\alpha)\}}$ 
        as a function of the rescaled coordinate, 
        $xt_*^{-1/(1-\omega)}$, 
        where $t_\mathrm{c}$ and $\eta$ are 
        set to 0.915 Myr and 0.98, respectively. The thick lines in (c) and (d) indicate the 
        corresponding S-S solution.
        }
        \label{inter}
\end{figure*}

Figure \ref{Pden evo} shows the evolutionary track of the gas
at the centre, $x=0$. 
Due to the shock compression, the temperature of gas 
increases suddenly, and the postshock region enters the thermally 
unstable phase from the initial stable 
state. The TI leads to the cooling layer condensing in a runaway fashion.
The cooling length in Fig. \ref{alpha evo} rapidly decreases with time until 
it reaches minimum value $1.5\times10^{-3}$ pc at $t=0.917$ Myr.
During runaway cooling, the gas condenses isobarically until it reaches 
the stable CNM phase, although some pressure oscillation is seen in Fig. \ref{Pden evo}.
We focus on the property of this runaway condensing layer. First 
we evaluate $\alpha_\mathrm{num}(t)$, which is defined by
\begin{equation}
        \alpha_\mathrm{num}=\left( \frac{ \partial (\rho{\cal L})}{\partial T} \right)
        _\mathrm{\rho}\Biggr|_{x=0}
\end{equation}
at each instant of time.
Figure \ref{alpha evo} shows the time evolution of $\alpha_\mathrm{num}$.
During $0.4<t/\mathrm{Myr}<0.9$, 
it is seen that $\alpha_\mathrm{num}\sim0.61$ is approximately constant.
Therefore, in this period, the flow is expected to be approximated well by the 
S-S solutions.

Figures \ref{inter} show
the time evolutions of the distributions of (a) the number density and (b) the temperature 
at $t=$ 0.72, 0.80, 0.88, and 0.90 Myr, respectively.
The rescaled number density 
$nt_*^{\eta/(2-\alpha)}$ 
and temperature, $T t_*^{\left\{ \eta - (2-\alpha) \right\}/\{(2-\alpha)(1-\alpha)\}}$
are shown in 
Figs. \ref{inter}c and \ref{inter}d 
as a function of rescaled coordinate, $xt_*^{-1/(1-\omega)}$, 
where $t_\mathrm{c}$, $\eta$, and $\alpha$
are set to 0.915 Myr, 0.98, and 0.61, respectively.
From Figs. \ref{inter}c and \ref{inter}d, it is clearly seen that the 
S-S solution $(\alpha=0.61, \eta=0.98)$ describes 
the results of the numerical calculation well.
There are two reasons they are described by the S-S solution well.
One is that $\alpha_\mathrm{num}$ is almost constant, and 
the cooling function is approximated well by Eq. (\ref{cooling func}) during the runaway condensation.
The other reason relates to the stability of S-S solutions.
The one-dimensional calculation solves the evolution only in the direction parallel to the condensation.
From the linear analysis in Sect. \ref{keq0}, 
such a perturbation with $k=0$ cannot grow enough during the 
runaway condensation.
That is why the S-S solution is realized even though the density is initially fluctuated.

However, the result is expected to be qualitatively different if the perturbation with $k\ne0$ exists. 
The perturbation with $k\ne0$ on the isobarically runaway condensing layer
grows as $\propto t_*^{-1}$ (see Sect. \ref{sec kne0}). The growth rate is independent of wave numbers.
From Figs. \ref{alpha evo} and \ref{inter}, 
the gas evolves obeying the S-S solution between $t=$ 0.4 and 0.9 Myr.
If the layer has a perturbation ($k\ne0$) with
amplitude $\delta \rho_0$ at $t=0.4\;\mathrm{Myr}$,
the perturbation grows as
\begin{equation}
        \frac{\delta \rho(t)}{\delta \rho_0}=\frac{1-0.4 \mathrm{Myr}/t_\mathrm{c}}
        {1 - t/t_\mathrm{c}}.
\end{equation}
The epochs when the perturbation grows by a factor of 10 and 100 are $t=$ 0.86 and 
0.91 Myr, respectively.
At 0.86 Myr, the central number density is still as low as 38 cm$^{-3}$.
Therefore, the condensing layer is expected to fragment quickly, and CNM clumps will form.

\cite{II08} investigated TI in a shock compressed region formed by WNM-WNM collision using 
two-dimensional, two-fluid magnetohydrodynamical simulation.
The initial condition there is the same as that in our one-dimensional simulation 
except for the dimension and the way to add density fluctuation in the WNM.
Our linear analysis is applicable to the gas evolution before
CNM clouds form in their unmagnetized case.
In the initial stage of their simulation, in the postshock region, 
thermally unstable gas initially condenses in a layer
structure. Around $t=0.5$ Myr when the region of the highest density reaches 
CNM stable state, both small CNM cloudlets 
and filamentary CNM clouds
are generated \citep{I09}. 
According to our linear analysis, an isobarically condensing gas layer 
is expected to 
split into fragments that have a variety of lengths in the 
transverse direction.
Therefore, our linear analysis agrees with their simulation qualitatively.

After the first CNM clouds formation, 
the mass of these CNM clouds continues to increase by the accretion 
of unstable gas by the shocks.
Moreover, some CNM clouds coalesce into larger clouds and
others fragment into smaller ones by the turbulent flow \citep{I09}. 
Therefore, the size and mass of CNM clouds varies with time.
To understand this phase, which is beyond the scope of this paper, 
a statistical approach is probably needed.
A statistical theory has been proposed by \cite{HA07}
using Press-Schechter formalism \citep{PS74}, assuming
that the CNM clumps are generated from density 
fluctuations within WNM whose
power spectrum is assumed to be Kolmogorov.

Comparing timescales, \cite{HHB08} discuss the effect of 
TI, the nonlinear thin-shell instability, Kelvin-Helmholtz instability,
and gravitational instability 
in various densities, temperatures, and scales of fluctuations.
In their paper, it is assumed that the gas is 
unstable only if the scale of the
fluctuation is smaller than the sound crossing scale.
However, from our linear analysis, 
perturbations can grow regardless of their scales
as long as they are flat rather than spherical.

\subsection{The growth rate for $1<\alpha<2$}
Although the linear analysis on the S-S cooling layer is limited for $\alpha<1$,
the thermal stability of the gas for $\alpha>1$ is also roughly understood from Balbus's criterion.
For $1<\alpha<2$, the gas  is isobarically unstable, but it is isochorically stable.
For $\alpha>2$, the gas  is thermally stable.
In this section, we investigate the stability of the gas for $1<\alpha<2$ during cooling 
within the large and small scale limits.

\subsubsection{The isobaric mode}
For the case with small wave length, perturbation is expected to grow isobarically. 
By comparison of our results with previous studies in the literature, it is found that 
the growth rate in the isobaric mode is independent of the global structure of 
the unperturbed state. 
Therefore, from local arguments, the growth rate in the isobaric mode of the gas with $1<\alpha<2$
can also be estimated.

As an unperturbed state, we adopt a cooling gas element whose
scale is assumed to be much smaller than the cooling length.
In this case, the element cools isobarically.
From Eq. (\ref{eoe}), the time evolution of the unperturbed gas
is given by 
\begin{equation}
 \rho(t)=\rho_i\left( 1-\frac{t}{t_\mathrm{cool}'} \right)^{-1/(2-\alpha)},
        \frac{1}{t_\mathrm{cool}'} = (2-\alpha)\gamma^{\alpha-1}(\gamma-1)
        P_i^{\alpha-1}\rho_i^{2-\alpha},
\label{a ge 1}
\end{equation}
where $\rho_i$ and $P_i$ represent the initial density and pressure, respectively.
In the above unperturbed state, we consider the following isobaric perturbation:
\begin{equation}
        \rho = \rho_0(t) +\delta \rho(t),
\end{equation}
and
\begin{equation}
        P = P_0,
\end{equation}
where subscript ``0'' indicates the unperturbed state, and 
$\delta \rho$ is the density perturbation.
Linearizing Eq. (\ref{eoe}), one obtains
\begin{equation}
   \frac{\mathrm{d} }{\mathrm{d}t}\left( \frac{\delta\rho}{\rho_0} \right)
 = (2-\alpha)\gamma\gamma^{\alpha-1}
 \rho_0^{2-\alpha}P_0^{\alpha-1}\frac{\delta \rho}{\rho_0}.
 \label{per a ge 1}
\end{equation}
Using Eq. (\ref{a ge 1}), Eq. (\ref{per a ge 1}) is rewritten as 
\begin{equation}
   \frac{\mathrm{d} }{\mathrm{d}t}\left( \frac{\delta\rho}{\rho_0} \right)
   = \frac{1}{t_\mathrm{cool}'} \left( 1-\frac{t}{t_\mathrm{cool}'} \right)^{-1}
   \frac{\delta \rho}{\rho_0}.
 \label{per a ge 1 2}
\end{equation}
Equation (\ref{per a ge 1 2}) is easily integrated to give 
\begin{equation}
        \frac{\delta \rho}{\rho_0} \propto \left( 1-\frac{t}{t_\mathrm{cool}'} \right)^{-1}.
        \label{a ge 1 growth rate}
\end{equation}
Comparing Eq. (\ref{a ge 1 growth rate}) with Eq. (\ref{a ge 1}),
one can see that the perturbation grows more slowly than the unperturbed state for $1<\alpha<2$.
Therefore, the gas is expected to be difficult to fragment during 
runaway cooling if $1<\alpha<2$.

\subsubsection{The noninteractive mode}
A cooling layer that evolves isobarically is considered.
The time evolution of the central density is the same as Eq. (\ref{a ge 1}).
When the scale of perturbation perpendicular to the condensation 
is too large to interact with other regions,
each region evolves independently.
Here, we focus on the time evolution of density perturbation at the centre ($x=0$).
Initial fluctuation of the central density, $\delta \rho_i$, 
creates the fluctuation of the cooling time, $\Delta t$.
The relative amplitude of density perturbation at $x=0$ is given by
\begin{equation}
        \frac{\delta \rho}{\rho_0} = \frac{1}{\rho_0}\left( \rho_i + \delta \rho_i \right)
        \left( 1-\frac{t}{t_\mathrm{cool}' - \Delta t}  
        \right)^{-1/(2-\alpha)}-1.
        \label{non int a ge 1}
\end{equation}
Linearizing Eq. (\ref{non int a ge 1}) with omitting terms that do not grow, we have 
\begin{equation}
\frac{\delta \rho}{\rho_0} = \frac{1}{2-\alpha}\frac{\Delta t}{t_\mathrm{cool}'} 
\left( 1-\frac{t}{t_\mathrm{cool}'} \right)^{-1}.
\label{non in a ge 1 1}
\end{equation}
Comparing Eq. (\ref{non in a ge 1 1}) with Eq. (\ref{a ge 1}),
we can see that the perturbation grows more slowly than the unperturbed state 
for $1<\alpha<2$.
Therefore, the gas is expected to be difficult to fragment for $1<\alpha<2$ for the 
large-scale perturbation, as well as the small scale.

\subsection{Effects of thermal conduction}\label{thermal conduction}
In this paper, the thermal conduction is neglected for simplicity.
However, for large wave number, the thermal conduction is 
expected to stabilize TI in the cooling layer \citep{F65}.
Therefore, there is a critical wave number, $k_\mathrm{crit}$, such that perturbation 
with a larger wave number is stabilized by the thermal conduction.

First, we evaluate $k_\mathrm{crit}$ using an order estimation.
Using the characteristic time scale of the thermal conduction, $t_\mathrm{diff}$, 
the diffusion equation is given by
\begin{equation}
        \frac{1}{t_\mathrm{diff}} \left( \frac{P_{00}}{\gamma-1} \right) \sim k^2K(T_{00}) T_{00},
\label{diffusion eq}
\end{equation}
where $K$ is the thermal conduction coefficient. From Eq. (\ref{diffusion eq}), 
the diffusion timescale is given by
\begin{equation}
        t_\mathrm{diff} \simeq 
        \frac{P_{00}}{(\gamma-1)K(T_{00})T_{00} } k^{-2}.
\label{diffusion}
\end{equation}
From Eq. (\ref{diffusion}), one can see that the diffusion timescale
is small for large wave number.
If $t_\mathrm{diff}<t_\mathrm{cool}$, the thermal conduction is expected to 
stabilize TI. Therefore, $k_\mathrm{crit}$ 
can be derived on the condition $t_\mathrm{diff}\sim t_\mathrm{cool}$ as
\begin{equation}
        k_\mathrm{crit} = \sqrt{\frac{k_\mathrm{K}}{L_\mathrm{cool}}},\;\;\mathrm{where}\;
        k_\mathrm{K} = \frac{P_{00}c_{00}}{(\gamma-1)KT_{00}}.
        \label{critical wave}
\end{equation}
\cite{F65} derived similar critical wave number to Eq. (\ref{critical wave}). 
Since the unperturbed state is time dependent, $k_\mathrm{crit}$ also 
evolves with time. 
Detailed evolution of $k_\mathrm{crit}$ depends on $K$.
In $T<6000$, we adopt $K=2.5\times10^3\sqrt{T}$ ergs cm$^{-1}$ K$^{-1}$ s$^{-1}$
\citep{P53}. 
In this case, from Eq. (\ref{critical wave}), 
the time evolution of $k_\mathrm{crit}$ can be derived analytically as
\begin{eqnarray}
\frac{\mathrm{d} \ln k_\mathrm{crit}}{\mathrm{d}\ln t_*}
&=& \frac{(2\alpha-1)\eta - (2-\alpha)(3-2\alpha)}{4(2-\alpha)(1-\alpha)}\nonumber\\
&=&\left\{\begin{array}{cc}
        \displaystyle - \frac{7-2\alpha}{4(2-\alpha)}<0 & \mathrm{for}\;\eta=1\\ 
        \displaystyle - \frac{3-2\alpha}{4(1-\alpha)}<0 & \mathrm{for}\;\eta=0\\ 
 \end{array}\right..
 \label{thermal cond}
\end{eqnarray}
For $\eta=1$ and $\eta=0$, it is found that Eq. (\ref{thermal cond}) is negative
for $\alpha<1$.
Since Eq. (\ref{thermal cond}) is the linear function for $\eta$,
$\mathrm{d}\ln k_\mathrm{crit}/\mathrm{d}\ln t_*$ is negative for all $\eta$.
Therefore, $k_\mathrm{crit}$ increases with time. This means that an initially 
stable perturbation with wave number ($k>k_\mathrm{crit}$) becomes unstable 
at a certain epoch when $k_\mathrm{cirt}$ catches up with $k$.

Figure \ref{alpha evo} shows the time evolution of 
the critical wave length, $\lambda_\mathrm{crit}=2\pi/k_\mathrm{crit}$.
In Sect. \ref{astro impli}, the time evolution of the condensing layer can be 
described by the S-S solution with ($\alpha_\mathrm{num}=0.61$, $\eta=0.98$).
In Fig. \ref{alpha evo}, we see that $\lambda_\mathrm{crit}$ decreases 
with time during the runaway condensation. 
Figure \ref{alpha evo} also shows that $\lambda_\mathrm{crit}<L_\mathrm{cool}$, or $\kappa_\mathrm{crit}=
k_\mathrm{crit}L_\mathrm{cool}>1$. 
This means that the effect of thermal conduction on the dispersion relation 
(Fig. \ref{kneq0 grow}) always appears in the isobaric regime during the runaway condensation.

\section{Summary}\label{summary}
In this paper, we have investigated the stability of S-S solutions describing the runaway 
cooling of a radiative gas by linear analysis.
The results of our investigation are summarized as follows, 

\begin{enumerate}
        \item For the case with perturbation only parallel to the flow ($k=0$), 
              the S-S solutions are unstable. However, the growth
              rate is too low to become nonlinear during the runaway cooling.
              Actually, the S-S solutions are realized in one-dimensional 
              hydrodynamical calculations in IT08 and 
              in Sect. \ref{astro impli} in this paper.

        \item For the case with transverse perturbation ($k\ne0$), 
              there are several unstable modes in the S-S solutions. 
              The most unstable modes are the isobaric mode for $k\gg1$ and 
              the noninteractive mode for $k\ll1$.
              In the isobaric mode, the perturbation grows in pressure 
              equilibrium with its surroundings.
              On the other hand, the noninteractive mode is originated from 
              each region in the layer condensing independently.
              Under a static approximation, we derive the approximated dispersion relation.
              The results of direct numerical integration of the time evolution 
              agree with those using the static approximation.

        \item The S-S solutions for $\eta > \alpha$ are unstable for any wavelength.
              Especially, if the unperturbed state is isobaric, the growth rate 
              is independent of wave number. Therefore, fluctuations in various scales 
              grow simultaneously, and the gas layer is expected to split 
              into fragments with various scales.
              The S-S solutions for $\eta < \alpha$ are only unstable in the 
              isobaric mode.
\end{enumerate}

Our linear analysis predicts that the cooling layer splits into fragments quickly even if 
the size is greater than the local cooling length. 
Our linear analysis is qualitatively
consistent with the results of recent multi-dimensional numerical simulations
until CNM clouds form, but
the evolution of CNM clouds is beyond the scope of this paper.
\section*{Acknowledgments}
We would like to thank Tsuyoshi Inoue for valuable discussions.
We also would like to thank the anonymous referee for valuable 
comments and suggestions that
improved this paper significantly.
K.I. is supported by grants-in-aid for JSPS Fellow (21-1979).


\begin{appendix} 
\section{Supplements for derivation of basic equations in the 
  zooming coordinate}\label{app S-S}
Here, we present preparation
for derivation of basic equations (\ref{sim1})-(\ref{sim3}) in the zooming
coordinate given by Eq. (\ref{zooming}). 
In the zooming coordinate, the physical variables, $Q(t,x,y)$,
are given by the following unified form:
\begin{equation}
      Q(t,x,y) = Q_0(t)\Theta(\tau,\xi,y),\;\;\;
      Q_0(t) \propto t_*^{-q/(1-\omega)},
\end{equation}
where $Q(t,x,y)$ corresponds to $[\rho, v_x, v_y, P]$ (see Eq. \ref{per vari}),
and $\Theta=[\Omega, V_x, V_y, \Pi]$ 
are the physical variables in the zooming coordinate.
From Eqs. (\ref{self vari1})-(\ref{self vari3}),
a parameter, $q$, is given by
\begin{equation}
        q = \left\{
        \begin{array}{ll}
                -\beta  &  \mathrm{for}\;\;Q=\rho \\
                -\omega & \mathrm{for}\;\;Q=v_x,\;v_y\\
                -2\omega -\beta& \mathrm{for}\;\;Q=P\\
        \end{array}
        \right..
\end{equation}
The temporal and spatial derivatives of $Q(t,x,y)$ in the ordinary coordinate 
can be expressed in the zooming coordinate as 
\begin{eqnarray}
   \left( \frac{\partial Q}{\partial t} \right)_{x,y}
   &=& \frac{\mathrm{d} Q_0(t)}{\mathrm{d} t} \Theta 
   + Q_0(t) \frac{\mathrm{d}\tau}{\mathrm{d}t}
        \frac{\partial \Theta}{\partial \tau}
        + Q_0(t) \left(\frac{\partial \xi}{\partial t}\right)_{x}
        \frac{\partial \Theta}{\partial \xi} \nonumber\\
        &=& \frac{v_0(t)}{x_0(t)}Q_0(t)
        \left[ q \Theta
               + \frac{\partial \Theta}{\partial \tau} 
               + \xi\frac{\partial \Theta}{\partial \xi} \right],
               \label{ope zoom1}
\end{eqnarray}
\begin{equation}
   \left( \frac{\partial Q}{\partial x} \right)_{t,y}
   = \frac{Q_0(t)}{x_0(t)} \frac{ \partial \Theta}{\partial \xi},\;\;
   \mathrm{and}\;\;
   \left( \frac{\partial Q}{\partial y} \right)_{t,x}
   = Q_0(t) \frac{ \partial \Theta}{\partial y},
               \label{ope zoom2}
\end{equation}
respectively, where 
we use 
\begin{equation}
        \frac{v_0(t)}{x_0(t)}=
        -\frac{\dot{x}_0(t)}{x_0(t)}=
        \frac{1}{(1-\omega)t_\mathrm{c}t_*}
\label{zoom rela1}
\end{equation}
from Eq. (\ref{self vari1}).      
Using Eqs. (\ref{ope zoom1}) and (\ref{ope zoom2}),
the Lagrangian time derivative of $Q$ is given by
\begin{eqnarray}
&& \hspace{-5mm}
   \left(\frac{\partial }{\partial t} + v_0(t) V_x \frac{\partial }{\partial x} 
          + v_0(t) V_y \frac{\partial }{\partial y} 
   \right)Q(t,x,y) \nonumber \\
   &&\hspace{-5mm}  = \frac{v_0(t)}{x_0(t)}Q_0(t)
        \left[  q 
               + \frac{\partial}{\partial \tau} 
               + \left( V_x + \xi \right)
               \frac{\partial }{\partial \xi} 
               + V_y x_0(t)\frac{\partial }{\partial y} \right]
               \Theta(\tau,\xi,y).
               \label{ope zoom3}
\end{eqnarray}
Using Eqs. (\ref{ope zoom1})-(\ref{ope zoom3}),
\begin{equation}
        \frac{v_0(t)}{x_0(t)} = \gamma^{\alpha-1} \Lambda_0 
        \rho_0(t)^{2-\alpha}P_0(t)^{\alpha-1},\;\;\mathrm{and}\;\;
        P_0(t)= \frac{\rho_0(t)v_0(t)^2}{\gamma},
\end{equation}
basic equations (\ref{sim1})-(\ref{sim3}) are derived
in the zooming coordinate.

   \section{Detailed expression of $A_{ik}$}\label{app per}
 In this appendix, we provide the detail expression of $A_{ik}$ as 
\begin{equation}
        A_{11}= -\left( \frac{V_0^2-X_0^2}{V_0} \right)\sigma  
         - \frac{X_0^2}{\gamma}(\ln\Pi_0)' 
         + \frac{X_0^2}{V_0}(\alpha-2)\epsilon_0 
\end{equation}
\begin{equation}
        A_{12}= -V_0(\ln \Omega_0)' + \sigma + \omega + V_0' 
        - \frac{X_0^2}{\gamma V_0}(\ln\Pi_0\Omega_0^{-\gamma})',
\end{equation}
\begin{equation}
        A_{13}= \frac{X_0^2}{\gamma}
        \left[ (\ln\Pi_0)' - \frac{\sigma + (\alpha-1)\gamma\epsilon_0}{V_0} \right],
\end{equation}
\begin{equation}
        A_{14}= - kx_0(t)V_0,
\end{equation}
\vspace{3mm}

\begin{equation}
        A_{21} = \frac{X_0^2}{\gamma}\left( 
         V_0(\ln\Pi_0)' - (\alpha-2)\gamma\epsilon_0 \right),
\end{equation}
\begin{equation}
        A_{22} = X_0^2(\ln \Omega_0)' - V_0(\sigma + \omega + V_0') 
        + \frac{X_0^2}{\gamma}(\ln\Pi_0\Omega_0^{-\gamma})',
\end{equation}
\begin{equation}
        A_{23}= -V_0A_{14},
\end{equation}
\begin{equation}
        A_{24}= kx_0(t)X_0^2,
\end{equation}
\vspace{3mm}
\begin{equation}
        A_{31} = - X_0^2(\ln\Pi_0)' + (\alpha-2)V_0\gamma\epsilon_0,
\end{equation}
\begin{equation}
        A_{32} = \gamma\left\{-V_0(\ln \Omega_0)' + \sigma + \omega + V_0'
        - \frac{V_0}{\gamma}(\ln \Pi_0\Omega_0^{-\gamma})\right\},
\end{equation}
\begin{equation}
        A_{33}= 
        X_0^2(\ln\Pi_0)' - V_0\{\sigma + (\alpha-1)\gamma\epsilon_0\},
\end{equation}
\begin{equation}
        A_{34}= - \gamma kx_0(t)V_0,
\end{equation}
\vspace{3mm}
\begin{equation}
        A_{41}= 0,
\end{equation}
\begin{equation}
        A_{42}= 0,
\end{equation}
\begin{equation}
        A_{43}= - \frac{V_0^2 - X_0^2}{V_0}(\omega + \sigma),
\end{equation}
and
\begin{equation}
        A_{44}= kx_0(t)(V_0^2-X_0^2)\frac{X_0^2}{\gamma}.
\end{equation}

\end{appendix}

\end{document}